\documentclass[11pt]{iopart}
\usepackage{graphicx,color,soul,amssymb,cite}

\usepackage{color}
\newcommand{\bn}{\bi{n}}
\newcommand{\bu}{\bi{u}}
\newcommand{\bz}{\mathbf{0}}
\newcommand{\bk}{\bi{k}}
\newcommand{\bfr}{\bi{r}}
\newcommand{\bv}{\bi{v}}
\newcommand{\D}{\Delta}

\let\OLDthebibliography\thebibliography
\renewcommand\thebibliography[1]{
  \OLDthebibliography{#1}
  \setlength{\parskip}{0pt}
  \setlength{\itemsep}{0pt plus 0.4ex}
}

\begin{document}
\title{Nature of self-diffusion in two-dimensional fluids}

\author{Bongsik Choi$^1$,
Kyeong Hwan Han$^1$, Changho Kim$^2$, Peter Talkner$^3$, Akinori Kidera$^4$, Eok Kyun Lee$^1$\footnote{corresponding author, eklee@kaist.ac.kr}}
\address{$^1$
 Department of Chemistry, Korea Advanced Institute of Science and Technology (KAIST), 291 Daehak-ro, Yuseong-gu, Daejeon 34141, Republic of Korea
}%
\address{$^2$
Computational Research Division, Lawrence Berkeley National Laboratory, 1 Cyclotron Road, Berkeley, California 94720, USA 
}%
\address{$^3$
Institut f\"ur Physik, Universit\"at Augsburg, Universit\"atsstra{\ss}e 1, 86159 Augsburg, Germany
}%
\address{$^4$
 Graduate School of Medical Life Science, Yokohama City University,  Yokohama 230-0045, Japan
}%


\date{\today}


\begin{abstract}

Self-diffusion in a two-dimensional simple fluid is investigated by both analytical and numerical means. We investigate the anomalous aspects of self-diffusion in two-dimensional fluids with regards to the mean square displacement, the time-dependent diffusion coefficient, and the velocity autocorrelation function using a consistency equation relating these quantities.
We numerically confirm the consistency equation by extensive molecular dynamics simulations for finite systems, corroborate earlier results indicating that the kinematic viscosity approaches a finite, non-vanishing value in the thermodynamic limit, and establish the finite size behavior of the diffusion coefficient. We obtain the exact solution of the consistency equation in the thermodynamic limit and use this solution to determine the large time asymptotics of the mean square displacement, the diffusion coefficient, and the velocity autocorrelation function. An asymptotic decay law of the velocity autocorrelation function resembles the previously known self-consistent form, $1/(t\sqrt{\ln t})$, however with a rescaled time.

\end{abstract}

\noindent{\it Keywords\/}: Self-diffusion, long-time tail, anomalous diffusion, molecular dynamics simulation, consistency equation 
\maketitle

\section{Introduction}
Self-diffusion is one of the basic transport mechanisms in liquids. Its theoretical investigation goes back to the pioneering work of Alder and Wainwright~\cite{Alder_1970,Wainwright_1971}. Using molecular dynamics (MD) simulations they found that the velocity autocorrelation function (VACF), which is defined as 
$C(t)\equiv\langle\bv(0)\cdot\bv(t)\rangle$, displays so-called long-time tails being characterized by an asymptotic decay proportional to $t^{-d/2}$. Here, the brackets denote a thermal equilibrium average and $d=2,3$ specifies the dimension of the space occupied by the considered liquid. Soon after, this finding was corroborated by Kawasaki~\cite{Kawasaki_1970} within mode-mode coupling theory. The long-time tails are caused by hydrodynamic interactions between a tagged particle and the vortex flow induced by the particle motion relative to the rest of the fluid. 
At the lower dimension $d =2$, the resulting $1/t$ long-time tail though is inconsistent within mode-coupling theory. A modified asymptotics based on a self-consistent argument was suggested in  \cite{Keyes_1975,Hoef_1991a,Hoef_1991b,Donev_2011} leading to a slightly faster decay according to
\begin{equation}
\label{VACF_SCMCT_tail}
C(t)\sim\frac{1}{t\sqrt{\ln{t}}}.
\end{equation}
The diffusion coefficient describing the self-diffusion of a fluid particle is defined as $D(t)\equiv\frac{1}{d}\int_0^t C(t') \mathrm{d}t'$ and consequently, according to (\ref{VACF_SCMCT_tail}), grows asymptotically in time as $\sqrt{\ln t}$. Yet another time-integration yields the mean square displacement (MSD) of a tagged particle, $\langle \D\bfr^2(t) \rangle = 2d\int_0^t D(t') \mathrm{d}t'$, where 
$\D\bfr(t)\equiv\bfr(t)-\bfr(0)$ denotes the spatial increment of the tagged particle in the time $t$. In the case of a two-dimensional fluid (\ref{VACF_SCMCT_tail}) asymptotically leads to $ \langle \D \bfr^2(t) \rangle \sim t \sqrt{\ln t}$.
There have been various studies of the actual asymptotic form of the VACF and  the diffusion coefficient as well as of the MSD by MD simulations \cite{Nie_2011,Ferrario_1997,Isobe_2008,Meier_2004,Dib_2006}, all of them, however, being inconclusive concerning the logarithmic corrections which turn out as too weak to be reliably detected. Most of these studies rather identified an algebraic behavior of the MSD of the form
\begin{equation}
\label{MSD_superdiffusion}
\langle\D \bfr^2(t)\rangle\sim t^\alpha.
\end{equation}
For $\alpha >1$ such a spreading is known as superdiffusion. This phenomenon was discovered more than 90 years ago for the separation of a pair of particles moving in a turbulent fluid \cite{Richardson_1926}. Superdiffusion has been extensively studied over the past thirty years or so and 
a multitude of possible mechanisms have been identified
\cite{Klafter_1987,Latora_1999,Metzler_2004,Shin_2010,Shin_2014,Metzler_2014,Zaburdaev_2015}. The probability density function (PDF) of the displacement $\D \bfr(t)$ provides a distinguishing criterion, which though does not uniquely identify the underlying mechanism. 
Gaussian PDFs are known for normal diffusion and were also found by Liu and Goree in a two-dimensional system of particles interacting via a Yukawa potential \cite{Liu_2007,Ott_2009}. On the other hand,  motions with a broad distribution of jumps or flights typically lead
to PDFs characterized by heavy tails~\cite {Zaburdaev_2015}.

The purpose of the present investigation is to resolve the puzzle of the anomalous behavior, both by analytical and numerical means. As the central result we find for the scaling exponent $\alpha$ the value $1$ of normal diffusion and identify a particular slowly varying function causing the MSD growing disproportionately in time. The analytical result is based on self-consistent mode-coupling theory \cite{Zwanzig_2001} and compared to numerical MD simulations. 

\section{Gaussian nature of displacement}
In the MD simulations we considered $N$ particles moving on a two-dimensional square with periodic boundary conditions under the influence of a pairwise Weeks--Chandler--Andersen potential \cite{Weeks_1971}. The number density was chosen as $\rho=0.6$ and the temperature as $T=1$. In this parameter region the superdiffusive behavior is most pronounced. The according Hamiltonian equations of motion were solved by means of a velocity Verlet algorithm with a time step $\Delta t = 10^{-3}$. For the precise specification of the used dimensionless units and for further details, see Shin et al.~\cite{Shin_2014}.

\begin{figure}\begin{center}
\includegraphics[width=13cm]{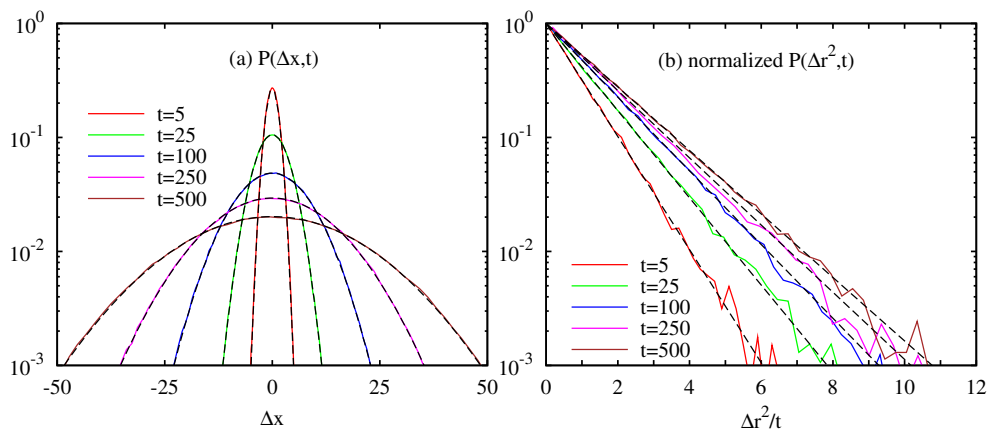} 
\caption{PDFs of $\Delta x$ obtained from MD simulations are presented in (a) for different times. The dashed lines represent Gaussian distributions with vanishing mean values and variances obtained from the MSD data.
In (b) the PDF of $\Delta r^2 \equiv \D \bfr^2(t)$ multiplied by the MSD $\langle \Delta \bfr^2(t) \rangle$ is displayed as a function of the squared displacement normalized by the time, $\Delta r^2/t$. The dashed lines represent exponential distributions $ P^{\mbox{exp}}(\D r^2,t) = \langle \D\bfr^2(t) \rangle^{-1} \exp \left \{ -\D r^2/ \langle \D\bfr^2(t) \rangle \right \}$.
}
\label{dspl_hist}
\end{center}\end{figure}
\begin{figure}\begin{center}
\includegraphics[width=13cm]{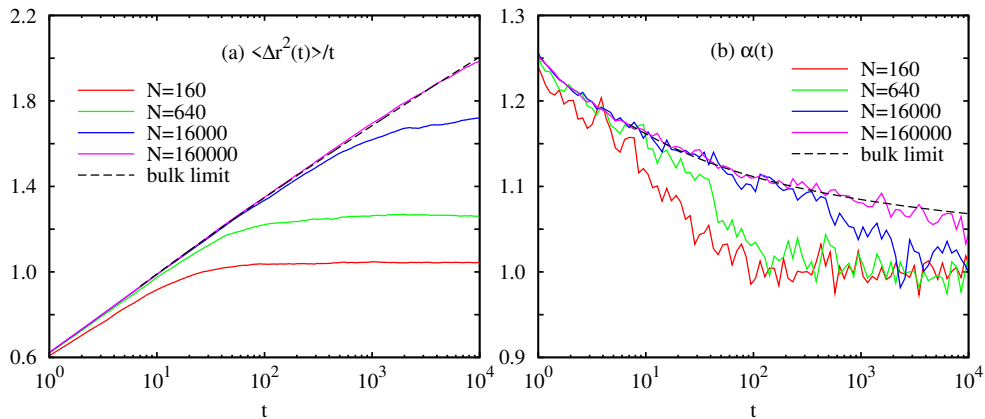}
\caption{The MSD per time, $\langle \Delta \bfr^2(t) \rangle /t$, is displayed in (a) as a function of time for several system sizes. The MSD follows the thermodynamic limiting law, (\ref{Ernst_ODE_solution_MSD}), up to a time depending on the size of the system and then approaches normal diffusive behavior. The local scaling exponent $\alpha(t)$ displayed in (b) decays on average monotonically until it reaches the value $1$ of normal diffusion. Black dashed lines in both panels were obtained from the analytical solution (\ref{Ernst_ODE_solution_MSD}).
}
\label{MSD_profile}
\end{center}\end{figure}

In figure~\ref{dspl_hist}(a) PDFs $P(\D{x})$ of the $x$-component of the displacement $\D \bfr (t)$, based on the corresponding histograms, are displayed semi-logarithmically for different times $t$. 
At all times the agreement with Gaussian distributions represented by a parabola is perfect. Also a Kolmogorov--Smirnov test confirms the Gaussian nature of the distributions at a high confidence level. Hence, for the considered times, the distribution of the displacement is fully characterized by its first, vanishing, moment and the second moment given by the MSD.
Deviations from a Gaussian displacement distribution might be expected only at the very short time scales characterizing the microscopic motion of the fluid particles.   
Figure~\ref{dspl_hist}(b) presents a semi-logarithmic plot of the PDFs of $\D r^2 \equiv \D \bfr^2(t)$ for different times $t$ as a function of $ \D r^2/t$. For normal diffusion all data would collapse  onto a single straight line. The decreasing inclination with increasing time however indicates a  superdiffusive spreading. In figure~\ref{MSD_profile}(a) this anomalous spreading is characterized by the MSD per time,  $\langle \D \bfr^2(t) \rangle / t$, for different system sizes $N$. As $N$ becomes larger, the disproportionate increase of the MSD with time becomes longer, however eventually it approaches normal diffusion motion growing proportionally to time. 
In order to quantify the increase of the MSD we determined a local exponent $\alpha(t)$ by subdividing time on a logarithmic scale into intervals of equal length $0.1$ on which the logarithm of the MSD was approximated by a linear function of $\ln t$. The local exponent was then estimated from the slope of the fitted straight line. Figure~\ref{MSD_profile}(b) exhibits the local exponent $\alpha(t)$  as, on average, decreasing with time until it eventually reaches the value $1$. The time at which the final value $\alpha =1$ is reached depends on the size $N$ of the system, which becomes larger as $N$ increases. As long as $\alpha(t)$ is larger than $1$, on average,
it is steadily decreasing  
without developing any plateau. This behavior presents a strong indication against a constant exponent $\alpha>1$.

\section{Self-consistency relation for finite size systems}
Following \cite{Ernst_1971,Erpenbeck_1982,Hansen_2013} we express the VACF at sufficiently large times by a sum over wave-vectors $\bk$ of the form
\begin{equation}
\label{finite_size_equation}
C(t)=\frac{k_\mathrm{B}T}{m\rho L^2}\sum_{\bk\neq\bz}\mathrm{e}^{-k^2 \int_0^t\{D(t')+\nu(t')\}\mathrm{d}t'},
\end{equation}
with $m$ denoting the particle mass and $L$ the side length of the system domain. Here the diffusion coefficient $D(t)$ and the kinematic viscosity coefficient $\nu(t)$ are determined by the respective Green--Kubo formulas $D(t) = \frac{1}{d}\int_0^t \mathrm{d}t' C(t')$ and $\nu(t) = \frac{L^2}{\rho k_\mathrm{B} T} \int_0^t \mathrm{d}t' \langle P_{xy}(0) P_{xy}(t') \rangle$, with $P_{xy}(t)$ denoting the off-diagonal element of the pressure tensor \cite{Rapaport_2004}. Details of the derivation of (\ref{finite_size_equation}) are given in the \ref{A}.

\begin{figure}\begin{center}
\includegraphics[width=8.25cm]{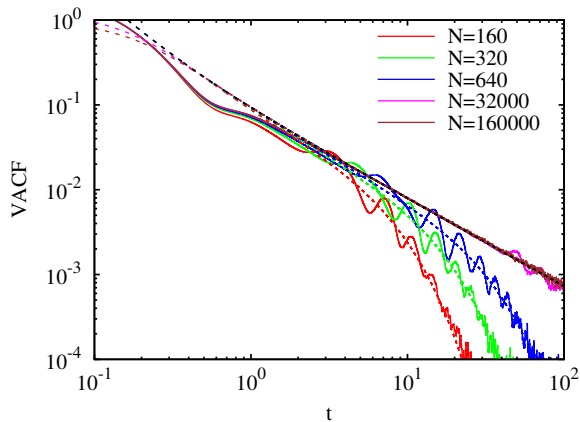} 
\caption{The graphs of the VACF resulting from MD simulations (solid line) and from the finite-size equation (\ref{finite_size_equation}) (dotted line) are compared with each other for different system sizes.}
\label{finite_size_eq}
\end{center}\end{figure}

We validated (\ref{finite_size_equation}) by the comparison of the MD results for the VACF and a numerical addition of the $\bk$-sum. The MD simulations were run for differently large systems with particle numbers $N=160,\;320,\;640,\;32\,000$ and $160\,000$ at the fixed density $\rho=0.6$. The results of this comparison are displayed in figure~\ref{finite_size_eq}.
The agreement is excellent for all but the smallest times at which the dynamics is still dominated by molecular kinetics rather than by hydrodynamic laws and also for those times $t=\sqrt{n^2+l^2} L/c_s$, with  integer numbers $n,l$, at which a signal propagating with sound velocity $c_\mathrm{s}$ may return to its starting point. Here the sound velocity is given by $c_\mathrm{s} =\sqrt{(\partial p/\partial \rho)_S/m}$ with the pressure $p$ and entropy $S$ \cite{Lustig_1994}. The small humps of $C(t)$ at the respective times  are not reproduced by the expression~(\ref{finite_size_equation}), because only the contribution of the diffusive transversal velocity field is considered and the propagation of longitudinal part is neglected~\cite{Hansen_2013}.
     
For finite systems both the viscosity coefficient and the diffusion coefficient converge to a finite value however with different size dependencies. While the diffusion constant $D=\lim_{t \to \infty} D(t) \propto \ln L$ diverges with  $L=\sqrt{N/\rho}$, the viscosity attains a finite value in this limit in accordance with other numerical studies of two-dimensional fluids with pairwise short-range repulsive interactions~\cite{Isobe_2009,Isobe_2010}. Based on MD simulations for different system sizes, we found $\nu =\lim_{t\to \infty} \nu(t) = 1.4769 - 0.9859/L$. For later use we note that the time-integral of the viscosity can be represented as $\int_0^t \mathrm{d}t' \nu(t') = \nu t +b$ where $b= -0.2674$ is almost independent of $L$. Further details of our study related to the viscosity will be published elsewhere.
\section{Self-consistency for infinitely large systems} 
In the thermodynamic limit, i.e. for $L \to \infty$ with the density $\rho$ kept constant, the sum on $\bk$ in (\ref{finite_size_equation}) can be replaced by an integral yielding 
\begin{equation}\label{Ernst_equation_generalized}
C(t)=\frac{k_\mathrm{B}T}{m\rho}\frac{1}{4\pi\int^{t}_0{\{D\left({t}^{'}\right)+\nu\left({t}^{'}\right)\}\mathrm{d}{t}^{'}}}.
\end{equation}
For three-dimensional systems the right-hand side is multiplied by a factor of two, and the denominator containing the time integral is taken to the power $\frac{3}{2}$. The diffusion as well as the viscosity coefficients can be replaced by constant values $D$ and $\nu$, respectively, leading to the notorious long-time tail of the VACF,  $C(t) \propto t^{-3/2}$ in three dimensions~\cite{Ernst_1971,Dorfman_1972,Pomeau_1975,Dib_2006}.

\begin{figure}[h]
\includegraphics[width=5.3cm]{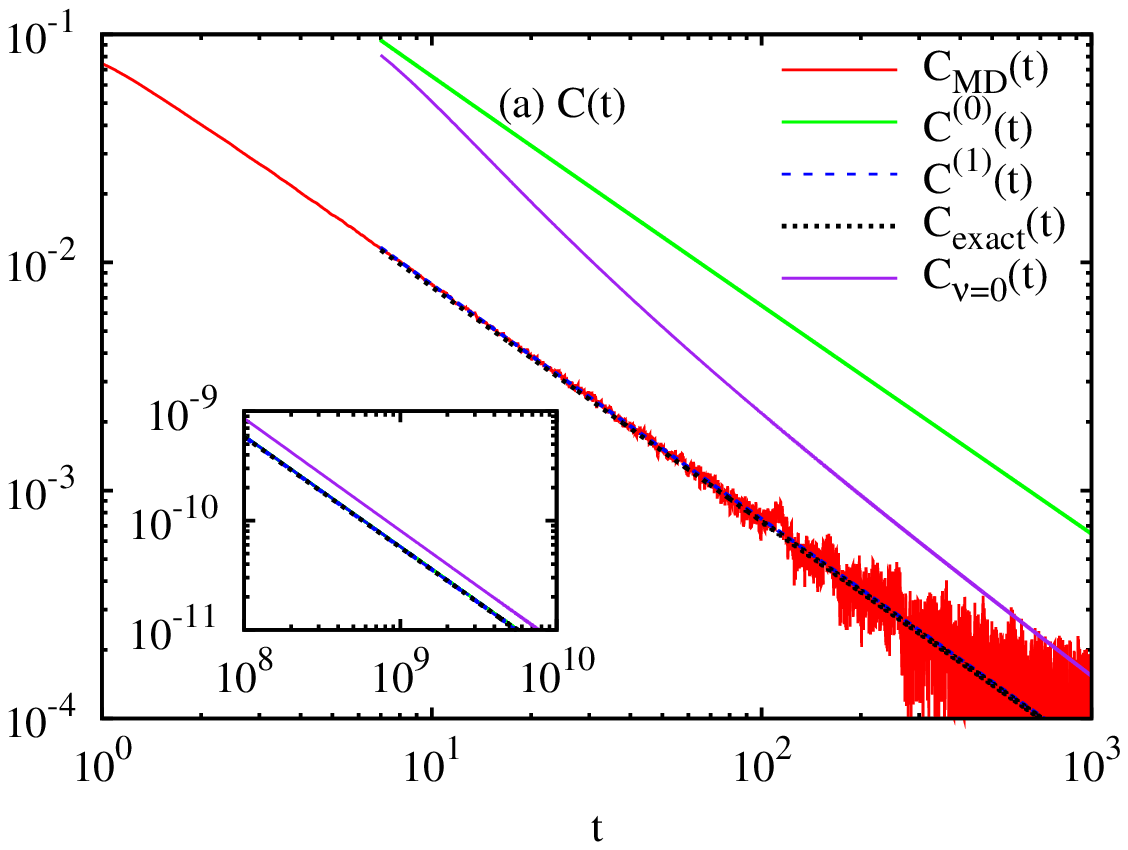} 
\includegraphics[width=5.3cm]{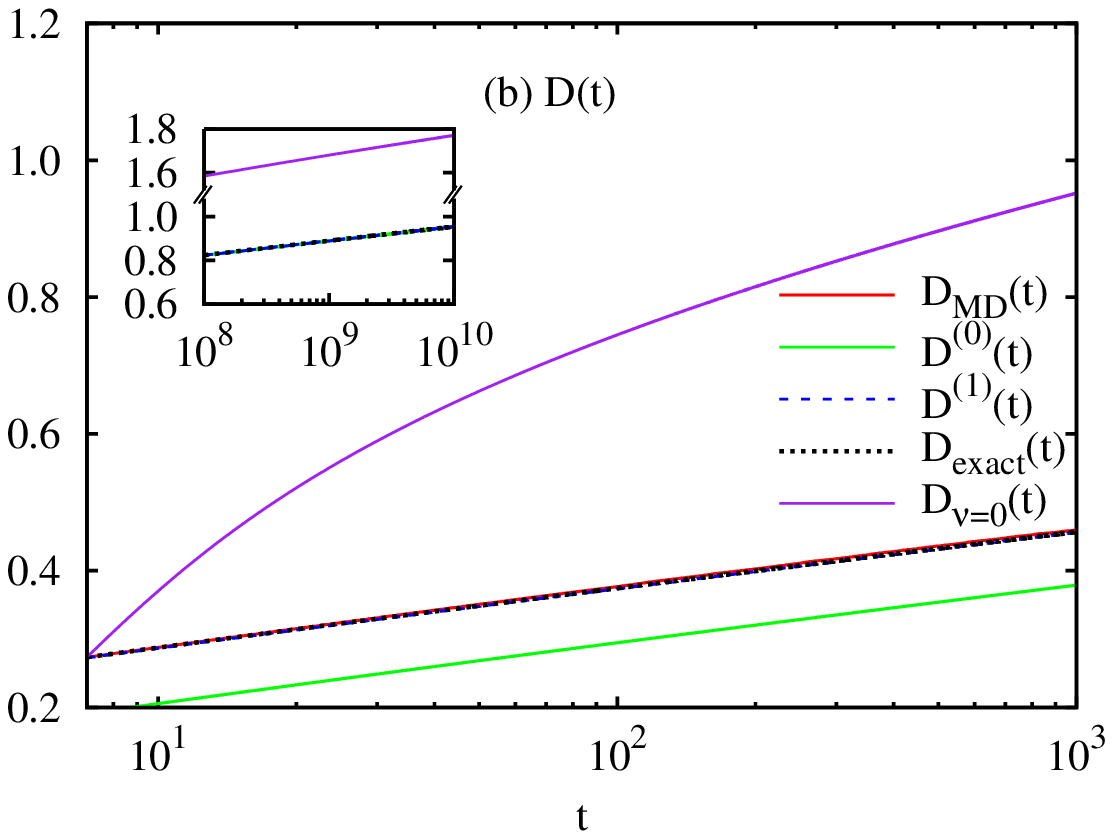} 
\includegraphics[width=5.3cm]{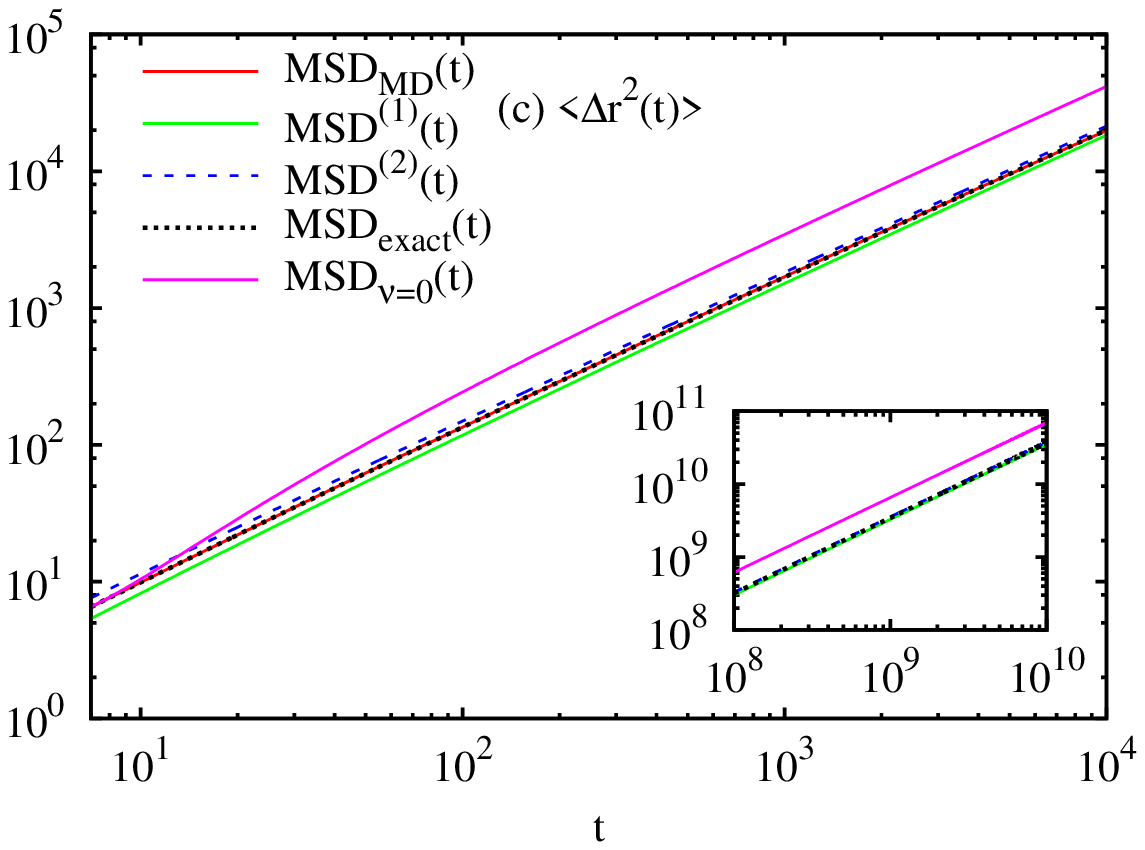} 
\caption{MD results for the VACF of a system with $160\,000$ particles as a function of time are compared in (a) with the exact expression~(\ref{Ernst_ODE_solution_C}) and several approximations thereof. The lowest-order approximation $C^{(0)}(t)$ and $C_{\nu=0}(t)$ show sensible deviations, while the first-order approximation $C^{(1)}(t)$ is virtually indistinguishable from the exact solution $C_\mathrm{exact}(t)$.
Likewise, the different levels of approximations for $D(t)$ and $\langle \D \bfr^2(t) \rangle$ are presented in panels (b) and (c), respectively. For $D(t)$ the exact expression~(\ref{Ernst_ODE_solution_D}), its approximation $D^{(1)}(t)$, and the MD simulation result agree with one another while $D^{(0)}(t)$ and $D_{\nu=0}(t)$ visibly deviate. In the case of MSD, the exact expression (\ref{Ernst_ODE_solution_MSD}), $\langle \D \bfr^2(t) \rangle^{(2)}$, and the MD result agree with each other, however $\langle \D \bfr^2(t) \rangle^{(1)}$ and $\langle \D \bfr^2(t) \rangle_{\nu = 0}$ show visible deviations. The lowest-order approximation $\langle \D\bfr^2(t) \rangle^{(0)}$ is not shown because it gives negative values. The inset in each panel demonstrates that the deviations of the $\nu=0$ expressions are still present for very large times, whereas $C^{(0)}(t)$, $D^{(0)}(t)$, $\langle \D\bfr^2(t) \rangle^{(1)}$, and higher-order approximations agree with the corresponding exact solutions.
}   
\label{Ernst_ODE}
\end{figure}

Substituting the viscosity integral by the form $\nu t+b$ inferred from our MD simulations, (\ref{Ernst_equation_generalized}) becomes a relation between the VACF $C(t)$ and the diffusion coefficient $D(t)$ of a two-dimensional fluid.
Further, introducing the function 
\begin{equation}
G(t) \equiv \int_0^t D(t') \mathrm{d}t' + \nu t +b= \frac{1}{4} \langle \D \bfr^2(t) \rangle + \nu t +b
\label{Gt}
\end{equation}
 and observing that $C(t) =2 \dot{D}(t) = 2 \ddot{G}(t)$ one obtains from (\ref{Ernst_equation_generalized}) the following closed equation for the auxiliary function $G(t)$:
\begin{equation}
\ddot{G}(t) = \frac{a}{G(t)},~~(t \ge t_0)
\label{Ernst_ODE_eq}
\end{equation}    
with $a=k_B T/(8 \pi m \rho) \approx 0.0663$. The auxiliary function $G(t)$ can be interpreted as the position of a particle of mass $1$ moving in a repulsive logarithmic potential $-a \ln G$. Consequently, the ``energy'' $E = \dot{G}^2(t)/2 - a \ln G(t)$ is conserved and the general solution of (\ref{Ernst_ODE_eq}) is readily found as
\begin{equation}\label{Ernst_ODE_solution}
G(t)=\exp\Big\{-\frac{E}{a}+\mathrm{erfi}^{-1}(s(t))^2\Big\},
\end{equation}
where $\mbox{erfi}^{-1}(z)$ denotes the inverse function of the imaginary error function $\mbox{erfi}(z) = \frac{2}{\sqrt{\pi}} \int_0^z \mathrm{d}u \mathrm{e}^{u^2}$, and 
\begin{equation}
s(t) = \sqrt{2 a} \mathrm{e}^{E/a}(t-t_0) + \mbox{erfi}\Big(\frac{|\dot{G}(t_0)|}{\sqrt{2 a}}\Big)
\label{st}
\end{equation}
is a scaled and shifted time variable. Using (\ref{Gt}) relating $G(t)$ and the MSD we determined the parameters $E= \dot{G}(t_0)^2/2 - a \ln G(t_0)$ and $c \equiv \mathrm{e}^{-E/a}\mbox{erfi} (|\dot{G}(t_0)|/\sqrt{2 a})/\sqrt{2 a} -   t_0$ for initial times varying in the interval $7~\le~t_0~\le~7.5$ in such a way that the difference between the MSD following from MD simulations of a system with $160\,000$ particles  and its value according to (\ref{Ernst_ODE_eq}) becomes minimal for times $7~\le~t~\le~110$. A comparison for larger times is not meaningful because of finite size effects manifesting themselves as a series of humps caused by the sound velocity of the fluid, see figure~\ref{finite_size_eq}.  We found optimal parameter values as $E=1.3688$ and $c=-0.1585$. These values turn out to be insensitive to the precise location of the time interval out of which the initial times $t_0$ is chosen. 
The MSD, the diffusion coefficient, and the VACF then assume the following forms
\begin{equation}\label{Ernst_ODE_solution_MSD}
\langle\D\bfr^2(t)\rangle=4\Big\{\exp\Big[-\frac{E}{a}+\mathrm{erfi}^{-1}(s(t))^2\Big]-\nu t-b\Big\},
\end{equation}
\begin{equation}\label{Ernst_ODE_solution_D}
D(t)=\sqrt{2a}~\mathrm{erfi}^{-1}(s(t))-\nu,
\end{equation}
\begin{equation}\label{Ernst_ODE_solution_C}
C(t)=2a\exp\Big\{\frac{E}{a}-\mathrm{erfi}^{-1}(s(t))^2\Big\}.
\end{equation}
Figure~\ref{Ernst_ODE} presents a comparison of the MSD according to (\ref{Ernst_ODE_solution_MSD}) and estimates from MD simulations. The agreement is excellent up to a characteristic time beyond which finite size effects become influential. 

In order to elucidate the asymptotic behavior of the above expressions  we make use of the approximation 
\begin{equation}
\mbox{erfi}^{-1}(s) \approx\left \{ \ln s+ \frac{1}{2} \ln \left [\pi \left ( \ln s + \frac{1}{2} \ln \left\{ \pi \left [ \ln s + \dots \right ] \right\} \right ) \right ] \right \}^{1/2}
\label{erfii}
\end{equation}
holding for large arguments $s$. It follows from the leading term of the asymptotic expansion of the imaginary error function $\mbox{erfi}(x) \approx \mathrm{e}^{x^2}/(x \sqrt \pi)$ by successive inversion~\cite{Abramowitz_1964}. For $C(t)$, $D(t)$, and $\langle \D \bfr^2(t)\rangle$, the influence of the order of approximation for the inverse imaginary error function is compared in figure~\ref{Ernst_ODE}. Using the lowest-order approximations, $\mbox{erfi}^{-1}(s) \approx  \{\ln s \}^{1/2}$ and  $\mbox{erfi}^{-1}(s) \approx  \{\ln s + \frac{1}{2} \ln (\pi \ln s) \}^{1/2}$, one obtains $C^{(0)}(t) = 2 a \mathrm{e}^{E/a} s^{-1}$ and  $C^{(1)}(t) = 2 a \mathrm{e}^{E/a} \left (s(t) \sqrt{\pi \ln s(t)} \right )^{-1}$. Likewise one finds $D^{(0)}(t) = \sqrt{2 a} \sqrt{\ln s(t)} -\nu$ and $D^{(1)}(t) = \{ 2 a [ \ln s(t) + \frac{1}{2}\ln (\pi \ln s(t)) ] \}^{1/2} -\nu$. Going to the next higher order approximation  one obtains $C^{(2)}(t) =2 a \mathrm{e}^{E/a} \{s(t) [\pi(\ln s(t)+\frac{1}{2}\ln \{\pi\ln s(t)\})]^{1/2} \}^{-1}$ for the VACF and accordingly $D^{(2)}(t) = \{ 2 a [ \ln s(t) + \frac{1}{2}\ln (\pi \{\ln s(t)+\frac{1}{2}\ln [\pi\ln s(t)]\}) ] \}^{1/2} -\nu$ for the diffusion coefficient. 
For the MSD it turns out that the lowest-order approximation $\langle \D\bfr^2(t) \rangle^{(0)} = 4\{ \mathrm{e}^{-E/a} s(t) -\nu t -b\}$ is insufficient because it attains a negative value. To the first-order one finds $\langle \D\bfr^2(t) \rangle^{(1)} = 4 \{\mathrm{e}^{-E/a} s(t) \sqrt{\pi \ln s(t)}-\nu t -b\}$ and to the second order $ \langle \D\bfr^2(t) \rangle^{(2)} = 4 \{\mathrm{e}^{-E/a} s(t) [\pi( \ln s(t)+\frac{1}{2}\ln \{\pi \ln s(t)\})]^{1/2}-\nu t -b\} $.
As demonstrated in figure~\ref{Ernst_ODE}, the VACF is already well described by the first-order approximation. Only if the viscosity is disregarded, yielding $C_{\nu=0}(t)$ and $D_{\nu=0}(t)$, a marked deviation is noticeable. For the diffusion coefficient and similarly for the MSD the first- and second-order approximations almost exactly agree with the exact solution and the MD result. 

The decisive difference between standard self-consistent mode-coupling theory and the present self-consistent theory relies on the appearance of the modified time-like variable  $s(t)$. Due to the large factor $\mathrm{e}^{E/a} \approx 10^8$ multiplying $t$ in $s(t)$ the logarithmic corrections $\ln s(t) \approx E/a + \ln t$ are strongly enhanced by an additive constant that becomes negligible only at extremely large times $t \gg 10^8$. 

\section{Conclusions}
We determined the asymptotic behavior of self-diffusion in two-dimensional liquids based on the thermodynamic-limit form~(\ref{Ernst_equation_generalized}) of the self-consistency relation  (\ref{finite_size_equation}) relating the VACF and the diffusion coefficient under the assumption that the viscosity approaches a finite, non-vanishing value in the thermodynamic limit. The resulting behavior of the VACF assumes the same functional form of 
the standard self-consistent mode-coupling theory $C(t) \propto (t \sqrt{\ln t})^{-1}$ with the essential difference that the time  is scaled by a large factor. While the scaling behavior as predicted by the standard theory sets in only at unobservably large times, our expression for $C(t) \propto (s(t) \sqrt{\ln s(t)})^{-1}$ as well as the according expressions for the diffusion coefficient and the MSD hold for all those times that are larger than the kinetic time scale set by the molecular interactions.

\ack This work was supported by the U.S.\ Department of Energy, Office of Science, Office of Advanced Scientific Computing Research, Applied Mathematics Program under Contract No.~DE-AC02-05CH11231, and Korea Advanced Institute of Science and Technology (KAIST), College of Natural Science, Research Enhancement Support Program under Grant No.~A0702001005.
The computation was partially done at the supercomputer system of Graduated School of Medical Life Science, Yokohama City University.
C.K. would like to thank Xiantao Li for informative discussions.

\appendix
\section{Derivation of the self-consistency condition (\ref{finite_size_equation})}\label{A}
Following \cite{Ernst_1971,Erpenbeck_1982,Hansen_2013} we review the derivation of (\ref{finite_size_equation})  based on the laws of linearized  hydrodynamics~\cite{Hansen_2013,Boon_1980}. 
This equation establishes a self-consistency relation between the VACF $C(t)$ and the time-dependent diffusion coefficient $D(t)$ if   the time-dependent kinematic viscosity coefficient $\nu(t)$ is known.

By introducing the average $\langle \bv(t)|\bv_0 \rangle$ of the velocity $\bv$ of a tagged particle at time $t$, \textit{conditioned} on the particle's initial velocity $\bv_0$, one can express the VACF in terms of the average over all initial velocities as
\begin{equation}
\label{VACF}
C(t) = \big< \bv_0 \cdot \langle \bv(t)|\bv_0 \rangle \big>.
\end{equation}
Note that the outer average is defined by the Maxwell--Boltzmann distribution 
at the temperature of the fluid.

The conditional average of the tagged particle can be approximated by using a spatial average with the fluid velocity field $\bu( \bfr,t;\bv_0)$, which is the solution of the linearized Navier--Stokes equations with the initial condition $\bu(\bfr,0;\bv_0) = \frac{1}{\rho} \delta (\bfr) \bv_0$.
That is, the conditional average is expressed as
\begin{equation}
\langle \bv(t)|\bv_0 \rangle = \int \bu(\bfr,t;\bv_0) P(\bfr,t)\mathrm{d}\bfr,
\label{tv}
\end{equation}
where $P(\bfr,t)$ describes the spreading of the tagged particle in space
being determined by the mass diffusion equation
\begin{equation}
\label{diffusion_equation}
\frac{\partial P(\bfr,t)}{\partial t}=D(t)\nabla^2 P(\bfr,t) 
\end{equation}
with the initial condition $P(\bfr,0)=\delta(\bfr)$.
The velocity field $\bu(\bfr,t;\bv_0)$ can be split into longitudinal and transversal components, $\bu_{||}(\bfr,t;\bv_0)$
and $\bu_\bot(\bfr,t;\bv_0)$, respectively.
The longitudinal part describes sound propagation, which is fast. It leads to a rapidly decaying contribution to the correlation function and therefore can be neglected at large times.
The remaining transversal contribution is governed by the following vorticity diffusion equation 
\begin{equation}
\label{utr_diffusion_equation}
\frac{\partial}{\partial t}\bu_{\bot} (\bfr,t;\bv_0)=\nu(t)\nabla^2 \bu_{\bot}(\bfr,t;\bv_0). 
\end{equation}
Both diffusion equations (\ref{diffusion_equation}) and (\ref{utr_diffusion_equation}) are conveniently solved by means of a spatial Fourier transformation, yielding
\begin{eqnarray}
\label{diffusion_equation_solution}
&\tilde{P}(\bk,t)=\mathrm{e}^{-k^2\int_0^t D(t')\mathrm{d}t'}, \\ 
\label{utr_diffusion_equation_solution}
&\tilde{\bu}_{\bot}(\bk,t;\bv_0)=\frac{1}{\rho}\Big\{\bv_0-\frac{\bv_0 \cdot \bk}{k^2}\bk \Big\}\mathrm{e}^{-k^2\int_0^t \nu(t')\mathrm{d}t'}, 
\end{eqnarray}
where a tilde specifies the spatial Fourier transformation and $\bk = 2 \pi \bn/L$ is a vector in the reciprocal space with $\bn =(n_x,n_y)$ being a pair of integers. Parseval's theorem allows one to transform the spatial integral in (\ref{tv}) into a sum over all allowed reciprocal vectors, yielding upon averaging over $\bv_0$ the desired result, (\ref{finite_size_equation}).
A similar equation was also obtained by Erpenbeck and Wood~\cite{Erpenbeck_1982}.\\

\bibliographystyle{iopart-num}
\bibliography{ms}
\end{document}